\documentclass[11pt, a4paper]{article}
\usepackage{tikz}
\usetikzlibrary{snakes}
\usepackage[utf8]{inputenc} % Any characters can be typed directly from the keyboard, eg éçñ
\usepackage{textcomp} % provide lots of new symbols
\usepackage{flafter}  % Don't place floats before their definition

\usepackage{amsmath,amssymb}  % Better maths support & more symbols
\usepackage{bm}  % Define \bm{} to use bold math fonts
\usepackage{authblk}
\usepackage{memhfixc}  % remove conflict between the memoir class & hyperref
\usepackage{listings}
\newcommand{\be}{\begin{equation}}
\newcommand{\ee}{\end{equation}}

\def\XXint#1#2#3{{\setbox0=\hbox{$#1{#2#3}{\int}$}
\vcenter{\hbox{$#2#3$}}\kern-.5\wd0}}

\numberwithin{equation}{section}
\begin{document}
\title{Non-Markovian reversible diffusion-influenced \\ reactions in two dimensions}
\author{Thorsten Pr\"ustel} 
\author{Martin Meier-Schellersheim} 
\affil{Laboratory of Systems Biology\\National Institute of Allergy and Infectious Diseases\\National Institutes of Health}
\maketitle
\let\oldthefootnote\thefootnote 
\renewcommand{\thefootnote}{\fnsymbol{footnote}} 
\footnotetext[1]{Email: prustelt@niaid.nih.gov, mms@niaid.nih.gov} 
\let\thefootnote\oldthefootnote 
\abstract
{
We investigate the reversible diffusion-influenced reaction of an 
isolated pair in the presence of a non-Markovian generalization 
of the backreaction boundary condition in two space dimensions. Following 
earlier work by Agmon and Weiss, we consider residence time probability densities  
that decay slower than an exponential and that are characterized by a parameter $0<\sigma\leq 1$.
We calculate an exact expression for the probability $S(t\vert\ast)$ that the initially bound particle is unbound, 
which is valid for arbitrary $\sigma$ and for all times.
Furthermore, we derive an approximate solution for long times.
We show that the ultimate fate of the bound state is complete
dissociation, as in the 2D Markovian case. However, the limiting value is approached quite differently: 
Instead of a $\sim t^{-1}$ decay, we obtain $1-S(t\vert\ast)\sim t^{-\sigma}\ln t$.
}
\section{Introduction}
It has been shown that the ultimate fate of an isolated pair for the reversible diffusion-influenced reaction is independent from the dimensionality of space.
In fact, the ultimate escape probability is unity in one, two and three space dimensions \cite{Agmon:1984, Agmon:1990p10, kimShin:1999, TPMMS_2012JCP}. However, the dimensionality does have an influence on how the limiting value for $S(t\vert\ast)$, which denotes the probability that the initially bound particle is unbound,
is approached: Denoting the number of dimensions by $d$, the decay goes as $1-S(t\vert\ast) \sim t^{-d/2}$, 
which implies for instance considerable ramifications for the off-rate. This dependence from the dimension
can be traced back to the return to the origin probability of the underlying random walk. In one and two dimensions (2D), Brownian motion is recurrent, i.e. the return to
the origin happens with probability one, while in 3D it is transient, meaning that the non-return probability does not vanish. 

It is reasonable that, apart from the diffusive motion of the particles, the type of their interaction potentially influences the time-dependence of the probability $S(t\vert\ast)$ and, in particular, its long-time behavior.
Indeed, in most cases one studied the problem by tacitly assuming a Markovian residence time probability density for the bound state. However, Agmon and Weiss \cite{AgmonWeiss1989}
demonstrated that abandoning the Markovian assumption changes the time-dependence of $S(t\vert\ast)$ in a substantial way.
They found a phase transition as a function of the parameter $\sigma$, which is a measure for the deviation from the Markovian density (where $\sigma = 1$ corresponds to the Markovian limit).       
In one dimension, they found a first-order phase transition in which the long-time behavior of  $S(t\vert\ast)$ undergoes a discontinuous change at $\sigma = 1/2$. In contrast, in 3D
they observed a second-order phase transition at $\sigma = 1$ and the ultimate fate is again complete dissociation, independent from $\sigma$.

In what follows, we will extend the analysis presented in Ref.~\cite{AgmonWeiss1989} to the 2D case.
We will investigate non-Markovian dissociation for an isolated pair of reversibly binding particles 
which move in an infinitely extended two dimensional plane. 
The case of two dimensions is special, because it is the critical dimension with regard to recurrence and transience \cite{Toussaint_Wilczek_1983}. We will show whether even weak recurrence is sufficient to obtain a first-order phase transition.

We would like to point out that the consideration of non-Markovian dissociation is not only theoretically interesting. In particular 
in view of the intricate and heterogeneous composition of biological cell membranes \cite{bethani:2010}, generalizations of the Markovian behavior could potentially be significant for more realistic descriptions of the corresponding local biochemistries. 

The organization of the paper is as follows. First, we will introduce the general theoretical context, especially the Smoluchowski equation and backreaction boundary condition (BC).
Next, we will discuss memory effects and a non-Markovian generalization of the backreaction BC. In theses parts of our presentation, we shall closely follow Agmon and Weiss \cite{AgmonWeiss1989}.
Then, we will apply the general formalism to the 2D case. We will obtain an integral representation for $S(t\vert\ast)$ in the time domain, which is valid for all times. 
Furthermore, we will present an approximate solution for the long-time behavior of the probability $S(t\vert\ast)$. Finally, we discuss similarities and differences to the non-Markovian 1D and 3D cases.
\section{Smoluchowski equation and backreaction BC}
Solutions of the diffusion (Smoluchowski) equation \cite{Smoluchowski:1917, carslaw1986conduction, Rice:1985} can be employed to 
study diffusion-influenced reactions of an isolated pair. The different
types of chemical reactions are taken into account by imposing
certain boundary conditions at the encounter distance. Typically, one 
considers the following scenario \cite{Agmon:1984, Agmon:1990p10, kimShin:1999, TPMMS_2012JCP}. Two spherical (or rather disklike in 2D) 
particles $A$ and $B$ with diffusion constants $D_{A}$
and $D_{B}$, respectively, may associate when their separation equals the
encounter distance $a$ to form a bound molecule $AB$. 
Such a system may equivalently be described as the diffusion of a point-like particle with diffusion constant 
$D = D_{A}+D_{B}$ around a static disk with radius $a$.

The irreversible
association reaction is described by the radiation BC that is characterized
by an intrinsic association constant $\kappa_{a}$ \cite{CollinsKimball}. In the reversible case, 
the bound state may dissociate again to form an unbound pair $A+B$ 
and the radiation boundary condition has to be generalized correspondingly. The so-called backreaction BC \cite{Agmon:1984, Pines, PinesAgmonI, PinesAgmonII, Agmon:1990p10, kimShin:1999, TPMMS_2012JCP} has been 
used extensively to take into account reversible association-dissociation reactions, but in its conventional form it assumes  
a Markovian probability density
for the residence time \cite{AgmonWeiss1989}
\begin{equation}\label{psiPDFM}
\psi(t) = \kappa_{d}e^{-\kappa_{d}t},
\end{equation}
or, equivalently, in the Laplace domain
\begin{equation}\label{psiMLaplace}
\tilde{\psi}(s) = \frac{\kappa_{d}}{\kappa_{d} + s}.
\end{equation}
Eq.~\eqref{psiPDFM} corresponds to the situation where the bound molecule can be sufficiently 
well described by a single, discrete state. However, the bound
state is typically comprised of a large number of different internal states. 
It follows that the associated probability density $\psi(t)$ for residence time
in the set of all possible bound states is given by a superposition of exponentials, which correspond to the individual states
\begin{equation}\label{psiPDFNonM}
\psi(t) = \sum^{\infty}_{n=0}\varphi_{n}e^{-\lambda_{n}t}.
\end{equation} 
Only if the time scales set by the
$\lambda_{n}$ parameters are sufficiently separated, i.e. the decay of the ground state (=dissociation) is much slower 
than the decay of all other internal states, Eq.~\eqref{psiPDFNonM} can be replaced by Eq.~\eqref{psiPDFM} 
and the Markovian description is justified. However, if there is no sufficient separation of the involved time scales, the effective description of all internal states 
as one single state cannot be Markovian any more.
Furthermore, as already mentioned in the introduction, a non-Markovian deviation could be caused by a complex environment, for instance in biological cellular systems.  

To proceed we consider the probability density function (PDF) $p(r, t\vert \ast)$, which yields the probability to find the particle at a distance equal to $r$, 
given that it was initially in the bound state.
The diffusion (or Smoluchowski) equation in 2D \cite{AgmonWeiss1989, Agmon:1990p10} governs the time evolution of $p(r, t\vert \ast)$ 
\begin{equation}
\frac{\partial}{\partial t}p(r,t\vert \ast) = D\bigg(\frac{\partial^{2}}{\partial r^{2}}p(r,t\vert \ast) + 
\frac{1}{r}\frac{\partial}{\partial r}p(r,t\vert \ast)\bigg), \quad r \geq a.  
\end{equation}
The initial condition is
\begin{equation}
p(r, t=0\vert \ast) = 0, \quad r > a.
\end{equation}
The PDF we are interested in is only defined for $r \geq a > 0$ and one has to impose a BC for $r=a$ specifying the behavior 
at the encounter distance. 
The (Markovian) backreaction BC incorporates association and dissociation by relating the diffusional flux at the surface of the 
"interaction disc" and the probability $S(t\vert \ast)$ that the initially bound particle is unbound at time $t>0$ as follows
\cite{Agmon:1984, Agmon:1990p10, kimShin:1999}
\begin{eqnarray}\label{BRBC}
2\pi a D \frac{\partial}{\partial r}p(r, t\vert \ast)\vert_{r=a} = \kappa_{a}p(a, t\vert \ast) - \kappa_{d}[1-S(t\vert \ast)].
\end{eqnarray}
Note that in 2D the diffusional flux is given by
\begin{eqnarray}
-J(a\vert \ast) = 2\pi a D \frac{\partial}{\partial r}p(r, t\vert \ast)\vert_{r=a},
\end{eqnarray} 
and the probability $S(t\vert \ast)$ is defined by
\begin{eqnarray}\label{defS}
S(t\vert \ast)  & = & 2\pi\int^{\infty}_{a} p(r, t\vert \ast) r dr.
\end{eqnarray}
Integrating the diffusion equation over space yields
\begin{equation}
J(a\vert \ast) = \frac{\partial}{\partial t}S(t\vert \ast),
\end{equation}
which translates in the Laplace domain to
\begin{equation}
\tilde{J}(a, s\vert \ast) = s \tilde{S}(s\vert \ast),
\end{equation}
where we have used that $S(0\vert \ast)$ vanishes.
Hence, the BC Eq.~\eqref{BRBC} becomes in the Laplace domain \cite{AgmonWeiss1989}
\begin{equation}\label{defJLaplace}
\tilde{J}(a, s\vert \ast) = \frac{\kappa_{d}-s\kappa_{a} \tilde{p}(a, s\vert \ast)}{s+\kappa_{d}}.
\end{equation}
For our analysis, we will make use of the following central identity  \cite{AgmonWeiss1989} that relates in the Laplace domain the probability that the particle is not bound to the PDF $\tilde{p}_{\text{ref}}(a,s\vert a)$, which corresponds to nonreactive diffusion and solves the diffusion equation subject to a reflecting BC, $\tilde{J}(a, s\vert a) = 0 $,
\begin{equation}\label{S_pref}
s\tilde{S}(s\vert \ast) = \frac{\kappa_{d}}{\kappa_{d} + s[1+\kappa_{a}\tilde{p}_{\text{ref}}(a,s\vert a)]}.
\end{equation} 
A salient feature of Eq.~\eqref{S_pref} is that it is valid in any dimension, thereby allowing to find an expression for $\tilde{S}$ once $\tilde{p}_{\text{ref}}(a,s\vert a)$ 
is known. In 2D, one has \cite{TPMMS_2012JCP}
\begin{equation}\label{pref2D}
\tilde{p}_{\text{ref}}(a,s\vert a)=\frac{1}{2\pi D}\bigg[I_{0}(qa)K_{0}(qa)+K_{0}(qa)K_{0}(qa)\frac{I_{1}(qa)}{K_{1}(qa)}\bigg],
\end{equation} 
where $I_{0}(z), I_{1}(z), K_{z}(x), K_{1}(z)$ refer to the modified Bessel function of first and second kind and zero and first order, respectively \cite[Sect.~9.6]{abramowitz1964handbook}.
Using $I_{0}(z)K_{1}(z)+I_{1}(z)K_{0}(z) = z^{-1}$ and combining Eqs.~\eqref{S_pref} and \eqref{pref2D} we arrive at
\begin{equation}
s\tilde{S}(s\vert \ast) = \frac{\kappa_{d}qK_{1}(qa)}{qK_{1}(qa)(s+\kappa_{d})+\kappa_{a}/(2\pi a D)sK_{0}(qa)}.
\end{equation}
To find an expression for the probability that the particle is unbound in the time domain, we could invert the Laplace transform explicitly by calculating a Bromwich contour integral.
However, in the Markovian backreaction BC case, an expression has already been derived by using alternatively the Green's function $p(r,t\vert t_{0})$ of the diffusion equation subject 
to a backreaction BC.
The obtained result is \cite{TPMMS_2012JCP}
\begin{equation}\label{boundSurvival}
S(t\vert *) = 1 - 2\frac{\kappa_{a}\kappa_{d}}{\pi^{3}a^{2}D^{2}}  \int^{\infty}_{0}e^{-Dt x^{2}} \frac{1}{\alpha^{2}(x)+\beta^{2}(x)} \frac{1}{x}dx,
\end{equation} 
where 
\begin{eqnarray}
\alpha(x) &=& ( x^{2} - \kappa_{D})J_{1}(xa) + hxJ_{0}(xa),\label{defAlpha}\\
\beta(x) &=& ( x^{2} - \kappa_{D})Y_{1}(xa) + hxY_{0}(xa), \label{defBeta}\\
h &=& \frac{\kappa_{a}}{2\pi aD}, \\
\kappa_{D}&=&\frac{\kappa_{d}}{D}.
\end{eqnarray}
$J_{n}(z), Y_{n}(z)$ denote the Bessel functions of first and second kind, respectively \cite[Sect.~9.1]{abramowitz1964handbook}.

We now seek the non-Markovian generalization of Eq.~\eqref{boundSurvival}. To this end, we have to find the non-Markovian analogue of the term 
$\kappa_{d}[1-S(t\vert \ast)]$ in the BC Eq.~\eqref{BRBC}. 
\subsection{Non-Markovian backreaction BC}
In Ref.~\cite{AgmonWeiss1989} it was shown that the backreaction BC can also be written in terms of the probability density of the residence time as follows
\begin{equation}\label{defNonMJ}
J(a, t\vert \ast) = -\kappa_{a}p(a, t\vert \ast) + \psi(t) + \kappa_{a}\int^{t}_{0}p(a, \tau\vert \ast)\psi(t-\tau)d\tau. 
\end{equation}  
Application of the Laplace transform leads to
\begin{equation}\label{defNonMJLaplace}
\tilde{J}(a, s\vert \ast) = \tilde{\psi}(s)-\kappa_{a}[1 - \tilde{\psi}(s)] p(a, s\vert \ast). 
\end{equation}  
Similarly, the Laplace transform of $S(t\vert\ast)$ may be written as \cite{AgmonWeiss1989}
\begin{equation}\label{defNonMS}
s\tilde{S}(s\vert \ast) = \frac{\tilde{\psi}(s)}{1+\kappa_{a}[1 - \tilde{\psi}(s)] \tilde{p}_{\text{ref}}(a, s\vert a)}. 
\end{equation} 
Here, two remarks are in order. First, upon using the Markovian version of the residence time probability density Eq.~\eqref{psiMLaplace} in
 Eqs.~\eqref{defNonMJLaplace}, \eqref{defNonMS}, one recovers Eqs.~\eqref{defJLaplace} and \eqref{S_pref}. On the other hand, starting from Eqs.~\eqref{defJLaplace} and \eqref{S_pref}, one can simply substitute $\kappa_{d}$ by $s\tilde{\psi}(s)/[1-\tilde{\psi}(s)]$ \cite{Agmon:1990p10} to arrive at Eqs.~\eqref{defNonMJLaplace}, \eqref{defNonMS}.  

Next, we turn our attention to the concrete form of the residence time density $\psi(t)$. In Ref.~\cite{AgmonWeiss1989}, the following choice of densities was employed
\begin{equation}\label{defPSI}
\tilde{\psi}(s)=\frac{1}{1+(s\kappa^{-1})^{\sigma}}, \quad 0 < \sigma \leq 1.
\end{equation}
For a discussion and motivation for this particular form, we refer to \cite{AgmonWeiss1989}. Here, we only note that for $\sigma = 1$ and by identifying $\kappa$ with $\kappa_{d}$, one recovers Eq.~\eqref{psiMLaplace}.  However, for $\sigma < 1$, $\psi(t)$ decays slower than an 
exponential. 
\section{Non-Markovian survival probability in 2D}
\begin{figure}
\includegraphics[scale=0.4]{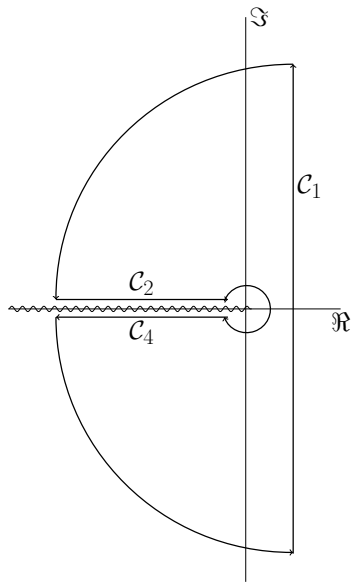}
\caption{Integration contour used  in Eq.~(\ref{cauchy}).}\label{fig:contour}
\end{figure}

In what follows, we will use Eqs.~\eqref{defNonMS}, \eqref{pref2D} and \eqref{defPSI} to derive an expression for $S(t\vert\ast)$ for arbitrary $0<\sigma\leq 1$ and all times.
We obtain
\begin{equation}\label{laplaceS}
\tilde{S}(s\vert \ast) = \frac{1}{s}\frac{\kappa^{\sigma}qK_{1}(qa)}{qK_{1}(qa)(s^{\sigma}+\kappa^{\sigma})+hs^{\sigma}K_{0}(qa)},
\end{equation}
where $q=\sqrt{s/D}$.
The inversion theorem for the Laplace transformation can be applied to find the corresponding expression of $\tilde{S}(s\vert\ast)$ in the time domain
\begin{equation}\label{inversionFormula}
S(t\vert\ast) = \frac{1}{2\pi i} \int^{\gamma+i\infty}_{\gamma-i\infty} e^{st}\,\tilde{S}(s\vert \ast )ds.
\end{equation}
To calculate the Bromwich contour integral, we first note that $\tilde{S}(s\vert\ast)$ has a branch point at $s=0$. Therefore, we use the contour of FIG.~\ref{fig:contour} with a branch cut along the negative real axis. We note that the contribution from the small circle around the origin does not vanish, but yields 1, which can be seen by the limiting forms of the modified Bessel functions for small arguments \cite[Eqs.~(9.6.7)-(9.6.9)]{abramowitz1964handbook}. Therefore, we obtain 
\begin{eqnarray}\label{cauchy}
\int^{\gamma+i\infty}_{\gamma-i\infty} e^{st}\,\tilde{S}(s\vert \ast )ds &=& 2\pi i -  \int_{\mathcal{C}_{2}} e^{st}\,\tilde{S}(s\vert \ast )ds  - \int_{\mathcal{C}_{4}} e^{st}\,\tilde{S}(s\vert \ast )ds.
\end{eqnarray}
It remains to calculate the integrals along $\mathcal{C}_{2}, \mathcal{C}_{4}$.
To this end, we choose
$
s = D x^{2} e^{i \pi }
$
and use \cite[Append.~3, Eqs.~(25), (26))]{carslaw1986conduction}
\begin{eqnarray}
I_{n}(xe^{\pm \pi i/2}) &=& e^{\pm n\pi i/2} J_{n}(x), \\
K_{n}(xe^{\pm \pi i/2}) &=& \pm\frac{1}{2}\pi i e^{\mp n\pi i/2} [-J_{n}(x) \pm i Y_{n}(x)].
\end{eqnarray}
It follows that
\begin{eqnarray}
\int_{\mathcal{C}_{2}}e^{st}\,\tilde{S}(s\vert \ast )ds  =  2\kappa^{\sigma}_{D}\int^{\infty}_{0}e^{-Dx^{2}t}\frac{[Y_{1}(xa)+iJ_{1}(xa)][\beta_{\sigma}(x)-i\alpha_{\sigma}(x)]}{\alpha_{\sigma}(x)^{2} + \beta_{\sigma}(x)^{2}}  \frac{dx}{x},
\end{eqnarray}
where $\kappa^{\sigma}_{D} = \kappa^{\sigma}/D^{\sigma}$ and we have defined 
\begin{eqnarray}
\alpha_{\sigma} &=& x^{2\sigma -1}\bigg[-\big( xJ_{1}(xa) + hJ_{0}(xa)\big)\cos(\pi\sigma)-\big( xY_{1}(xa) + hY_{0}(xa)\big)\sin(\pi\sigma)\bigg] \nonumber\\ 
&& - \kappa^{\sigma}_{D}J_{1}(xa), \\
\beta_{\sigma} &=& x^{2\sigma -1}\bigg[\big( xJ_{1}(xa) + hJ_{0}(xa)\big)\sin(\pi\sigma)-\big( xY_{1}(xa) + hY_{0}(xa)\big)\cos(\pi\sigma)\bigg] \nonumber\\
&&- \kappa^{\sigma}_{D}Y_{1}(xa).
\end{eqnarray}
We observe that $\alpha_{\sigma=1} = \alpha, \beta_{\sigma=1} = \beta$, cp. Eqs.~\eqref{defAlpha}, \eqref{defBeta}.

To evaluate the integral along the contour $\mathcal{C}_{4}$ we choose $p = Dx^{2}e^{-i\pi}$ and after an analogous calculation one finds that
$
\int_{\mathcal{C}_{2}}e^{st}\,\tilde{S}(s\vert\ast)ds=-\bigl(\int_{\mathcal{C}_{4}}e^{st}\,\tilde{S}(s\vert\ast)ds\bigr)^{\star},
$
where $\star$ means complex conjugation.
Thus, we have 
\begin{eqnarray}
S(t\vert\ast) - 1 = -\frac{1}{\pi} \Im\biggl(\int_{\mathcal{C}_{2}}e^{st}\,\tilde{S}(s\vert\ast)ds\biggr),
\end{eqnarray}
and arrive finally at the following expression for the probability to find the initially bound particle unbound in the time domain
\begin{eqnarray}\label{Sfinal}
S(t\vert\ast) = 1 + \frac{2}{\pi}\kappa^{\sigma}_{D}\bigg[\frac{2h}{a\pi}\cos(\pi\sigma)Q_{1} + \sin(\pi\sigma)Q_{2}\bigg], 
\end{eqnarray}
where
\begin{eqnarray}
Q_{1} &=& \int^{\infty}_{0}e^{-Dtx^{2}}\frac{x^{2\sigma-3}}{\alpha^{2}_{\sigma}(x)+\beta^{2}_{\sigma}(x)}dx,\\
Q_{2} &=& -\int^{\infty}_{0}e^{-Dtx^{2}}\frac{x^{2\sigma-2}\Omega(x)}{\alpha^{2}_{\sigma}(x)+\beta^{2}_{\sigma}(x)}dx,
\end{eqnarray}
and we have introduced
\begin{eqnarray}
\Omega(x) = x\big[J_{1}^{2}(xa)+Y_{1}^{2}(xa)\big]+h\big[J_{0}(xa)J_{1}(xa)+Y_{0}(xa)Y_{1}(xa)\big].
\end{eqnarray}
Eq.~\eqref{Sfinal} shows that the parameter $\sigma$ appears as a sort of mixing angle for the two qualitatively different contributions $Q_{1}, Q_{2}$.
We note that the term $Q_{2}$ is foreign to the Markovian expression Eq.~\eqref{boundSurvival}. Correspondingly, the second term vanishes for $\sigma=1$, while one recovers 
the Markovian limiting case Eq.~\eqref{boundSurvival} due to the first term $Q_{1}$ (we set $\kappa :=\kappa_{d})$.

Furthermore, we may already conclude from Eq.~\eqref{Sfinal} that in 2D the ultimate escape probability is unity for all $\sigma$, like in the Markovian and in the 3D non-Markovian case.
This behavior, however, is quite different to the non-Markovian 1D case, where the escape probability depends on the parameter $\sigma$ \cite{AgmonWeiss1989}. 

Finally, we are interested, how the limiting value is approached. To this end, we derive the asymptotic behavior for $t\rightarrow \infty$. Starting point is the 
expression for $\tilde{S}(s\vert\ast)$ Eq.~\eqref{laplaceS}. Using the series expansions of the modified Bessel functions for small arguments \cite[Eqs.~(9.6.10)-(9.6.13)]{abramowitz1964handbook},
we obtain
\begin{equation}
\tilde{S}(s\vert\ast) = \frac{1}{s} + \frac{ha}{\kappa^{\sigma}}s^{\sigma-1}\ln\bigg(\frac{1}{2}e^{\gamma_{E}}qa\bigg)  - \frac{s^{\sigma-1}}{\kappa^{\sigma}}+\ldots,
\end{equation}
where $\gamma_{E}$ denotes Euler's number $\gamma_{E}=0.5772156649\ldots$ \cite[Eqs.~(6.1.3)]{abramowitz1964handbook}.
For $\sigma = 1$ we obtain in the time domain
\begin{equation}
S(t\vert\ast) = 1 - \frac{\kappa_{a}}{\kappa_{d}}\frac{1}{4\pi Dt} + \ldots,
\end{equation}
which is the known Markovian result. If $ 0 < \sigma < 1$, we use \cite[Eqs.~(29.3.99), (29.3.7)]{abramowitz1964handbook}
\begin{eqnarray}
\mathcal{L}^{-1}\bigg[\frac{1}{s^{k}}\ln s\bigg] &=& \frac{t^{k-1}}{\Gamma(k)}\bigg[\Psi(k) - \ln t\bigg], \quad k > 0, \\
\mathcal{L}^{-1}\bigg[\frac{1}{s^{k}}\bigg] &=& \frac{t^{k-1}}{\Gamma(k)}, \quad k > 0,
\end{eqnarray}  
where $\mathcal{L}^{-1}, \Gamma, \Psi$ denote the inverse Laplace transform, the gamma function and the digamma function \cite[Eqs.~(6.3.1)]{abramowitz1964handbook}, respectively. 
Hence, we find 
\begin{equation}
S(t\vert\ast) = 1 - \frac{\kappa_{a}}{\kappa^{\sigma}}\frac{1}{4\pi D}\frac{t^{-\sigma}}{\Gamma(1-\sigma)}\ln t + C\frac{t^{-\sigma}}{\Gamma(1-\sigma)}+\ldots,
\end{equation}
where we introduced the constant
\begin{equation}
C=\frac{\kappa_{a}}{\kappa^{\sigma}}\frac{1}{2\pi D}\bigg[\frac{\Psi(1-\sigma)}{2}+\ln\bigg(\frac{1}{2}e^{\gamma_{E}}D^{-1/2}a\bigg)\bigg]-\frac{1}{\kappa^{\sigma}}.
\end{equation}
 
We find again that, independent from the parameter $\sigma$, the ultimate fate is complete dissociation, although the underlying random walk is recurrent. In contrast, in 1D the parameter $\sigma$ decides if particles escape or remain bound at long times. In this regard, the 2D case is more similar to the (non-Markovian) 3D case. However, it is very different to the 3D case with respect to how the limiting value is approached.
In 3D, it was shown that the decay goes as $t^{-\sigma}$ at $\sigma < 1$, but as $t^{-3/2}$ at $\sigma = 1$, which means there is a second-order phase transition. However, in 2D, even the functional form is different: While we observe a  $t^{-\sigma}\ln t$ decay at $\sigma < 1$, we find $t^{-1}$ at $\sigma = 1$. 
\subsection*{Acknowledgments}
This research was supported by the Intramural Research Program of the NIH, National Institute of Allergy and Infectious Diseases. 

We would like to thank Bastian R. Angermann and Frederick Klauschen for stimulating discussions.
\bibliographystyle{plain}

%\bibliography{Markovian} 
\end{document}